\documentclass[aps,prc,twocolumn,showpacs,superscriptaddress,amsmath,amssymb]{revtex4}

\usepackage{graphicx}
\usepackage{dcolumn}
\usepackage{bm}
\begin{document}

\preprint{APS/123-QED}

\title{The (t,$^{3}$He) and ($^{3}$He,t) reactions as probes of Gamow-Teller strength.}

\author{R.G.T. Zegers}
\affiliation{National Superconducting Cyclotron Laboratory, Michigan State University, East Lansing, MI 48824-1321, USA}
\affiliation{Department of Physics and Astronomy, Michigan State University, East Lansing, MI 48824, USA}
\affiliation{Joint Institute for Nuclear Astrophysics, Michigan State University, East Lansing, MI 48824, USA} 
\author{H. Akimune}
\affiliation{Department of Physics, Konan University, Kobe, Hyogo, 658-8501, Japan}
\author{Sam M. Austin}
\affiliation{National Superconducting Cyclotron Laboratory, Michigan State University, East Lansing, MI 48824-1321, USA}
\affiliation{Joint Institute for Nuclear Astrophysics, Michigan State University, East Lansing, MI 48824, USA} 
\author{D. Bazin}
\affiliation{National Superconducting Cyclotron Laboratory, Michigan State University, East Lansing, MI 48824-1321, USA}
\author{A.M. van den Berg}
\affiliation{Kernfysisch Versneller Instituut, University of Groningen, Zernikelaan 25, 9747 AA Groningen, The Netherlands}
\author{G.P.A. Berg}
\affiliation{Department of Physics, University of Notre Dame, IN 46556-5670, USA}
\affiliation{Joint Institute for Nuclear Astrophysics, University of Notre Dame, IN 46556-5670, USA}
\author{B.A. Brown}
\affiliation{National Superconducting Cyclotron Laboratory, Michigan State University, East Lansing, MI 48824-1321, USA}
\affiliation{Department of Physics and Astronomy, Michigan State University, East Lansing, MI 48824, USA}
\affiliation{Joint Institute for Nuclear Astrophysics, Michigan State University, East Lansing, MI 48824, USA} 
\author{J. Brown}
\affiliation{Department of Physics, Wabash College, Crawfordsville, IN 47933, USA}
\author{A.L. Cole}
\affiliation{National Superconducting Cyclotron Laboratory, Michigan State University, East Lansing, MI 48824-1321, USA}
\affiliation{Joint Institute for Nuclear Astrophysics, Michigan State University, East Lansing, MI 48824, USA} 
\author{I. Daito}
\affiliation{Advanced Photon Research Center, Japan Atomic Research Institute, Kizu, Kyoto 619-0215, Japan}
\author{Y. Fujita}
\affiliation{Department of Physics, Osaka University, Toyonaka, Osaka 560-0043, Japan}
\author{M. Fujiwara}
\affiliation{Kansai Photon Science Institute, Japan Atomic Energy Agency, Kizu, Kyoto 619-0215, Japan}
\affiliation{Research Center for Nuclear Physics, Osaka University, Ibaraki, Osaka 567-0047, Japan}
\author{S. Gal\`{e}s}
\affiliation{Institut de Physique Nucl\'{e}aire, IN2P3-CNRS, Orsay, France}
\author{M.N. Harakeh}
\affiliation{Kernfysisch Versneller Instituut, University of Groningen, Zernikelaan 25, 9747 AA Groningen, The Netherlands}%
\author{H. Hashimoto}
\affiliation{Research Center for Nuclear Physics, Osaka University, Ibaraki, Osaka 567-0047, Japan}
\author{R. Hayami}
\affiliation{Department of Physics, University of Tokushima, Tokushima 770-8502, Japan} 
\author{G.W. Hitt}
\affiliation{National Superconducting Cyclotron Laboratory, Michigan State University, East Lansing, MI 48824-1321, USA}
\affiliation{Department of Physics and Astronomy, Michigan State University, East Lansing, MI 48824, USA}
\author{M.E. Howard}
\affiliation{Joint Institute for Nuclear Astrophysics, Michigan State University, East Lansing, MI 48824, USA} 
\affiliation{Department of Physics, The Ohio State University, Columbus, OH 43210, USA}
\author{M. Itoh}
\affiliation{Cyclotron and Radioisotope Center, Tohoku University, Sendai, Miyagi 980-8578, Japan}
\author{J. J\"anecke}
\affiliation{Department of Physics, University of Michigan, Ann Arbor, MI 48109-1040, USA}
\author{T. Kawabata}
\affiliation{Center for Nuclear Study, University of Tokyo, RIKEN Campus, Wako, Saitama 351-0198, Japan}
\author{K. Kawase}
\affiliation{Research Center for Nuclear Physics, Osaka University, Ibaraki, Osaka 567-0047, Japan}
\author{M. Kinoshita}
\affiliation{Department of Physics, Konan University, 8-9-1 Okamoto Higashinda, Kobe, Hyogo, 658-8501, Japan}
\author{T. Nakamura}
\affiliation{Tokyo Institute of Technology, Megro, Tokyo 152-8550, Japan}
\author{K. Nakanishi}
\affiliation{Research Center for Nuclear Physics, Osaka University, Ibaraki, Osaka 567-0047, Japan}
\author{S. Nakayama}
\affiliation{Department of Physics, University of Tokushima, Tokushima 770-8502, Japan} 
\author{S. Okamura}
\affiliation{Research Center for Nuclear Physics, Osaka University, Ibaraki, Osaka 567-0047, Japan}
\author{W.A. Richter}
\affiliation{Department of Physics, University of Western Cape, Bellville 7530, South Africa}
\author{D.A. Roberts}
\affiliation{Department of Physics, University of Michigan, Ann Arbor, MI 48109-1040, USA}
\author{B.M. Sherrill}
\affiliation{National Superconducting Cyclotron Laboratory, Michigan State University, East Lansing, MI 48824-1321, USA}
\affiliation{Department of Physics and Astronomy, Michigan State University, East Lansing, MI 48824, USA}
\affiliation{Joint Institute for Nuclear Astrophysics, Michigan State University, East Lansing, MI 48824, USA} 
\author{Y. Shimbara}
\affiliation{National Superconducting Cyclotron Laboratory, Michigan State University, East Lansing, MI 48824-1321, USA}
\affiliation{Joint Institute for Nuclear Astrophysics, Michigan State University, East Lansing, MI 48824, USA} 
\author{M. Steiner}
\affiliation{National Superconducting Cyclotron Laboratory, Michigan State University, East Lansing, MI 48824-1321, USA}
\author{M. Uchida}
\affiliation{Tokyo Institute of Technology, 2-12-1 O-Okayama, Tokyo 152-8550, Japan}
\author{H. Ueno}
\affiliation{Applied Nuclear Physics Laboratory, RIKEN, Wako, Saitama 351-0198, Japan}
\author{T. Yamagata}
\affiliation{Department of Physics, Konan University, 8-9-1 Okamoto Higashinda, Kobe, Hyogo, 658-8501, Japan}
\author{M. Yosoi}
\affiliation{Research Center for Nuclear Physics, Osaka University, Ibaraki, Osaka 567-0047, Japan}
\date{\today}%

\begin{abstract}
Charge-exchange reactions are an important tool for determining weak-interaction rates. They provide stringent tests for nuclear structure models necessary for modeling astrophysical environments such as neutron stars and core-collapse supernovae. In anticipation of (t,$^{3}$He) experiments at 115 MeV/nucleon on nuclei of relevance (A$\sim40-120$) in the late evolution of stars, it is shown via a study of the $^{26}$Mg(t,$^{3}$He) reaction that this probe is an accurate tool for extracting Gamow-Teller transition strengths. To do so, the data are complemented by results from the $^{26}$Mg($^{3}$He,t) reaction at 140 MeV/nucleon which allows for a comparison of T=2 analog states excited via the mirror reactions. Extracted Gamow-Teller strengths from $^{26}$Mg(t,$^{3}$He) and $^{26}$Mg($^{3}$He,t) are compared with those from $^{26}$Mg(d,$^{2}$He) and $^{26}$Mg(p,n) studies, respectively. A good correspondence is found, indicating probe-independence of the strength extraction. Furthermore, we test shell-model calculations using the new USD-05B interaction in the sd-model space and show that it reproduces the experimental Gamow-Teller strength distributions well.
A second goal of this work is to improve the understanding of the (t,$^{3}$He) and ($^{3}$He,t) reaction mechanisms at intermediate energies since detailed studies are scarce. The Distorted-Wave Born Approximation is employed, taking into account the composite structures of the $^{3}$He and triton particles. The reaction model provides the means to explain systematic uncertainties at the 10-20\% level in the extraction of Gamow-Teller strengths as being due to interference between Gamow-Teller $\Delta L=0$, $\Delta S=1$ and $\Delta L=2$, $\Delta S=1$ amplitudes that both contribute to transitions from $0^{+}$ to $1^{+}$ states.  
 
\end{abstract}

\maketitle

\section{Introduction}
\label{sec:intro}

Charge-exchange reactions have long been used to study the spin-isospin response in nuclei. They transform a neutron (proton) into a proton (neutron) (change in isospin $\Delta$T=1), either with or without spin-transfer ($\Delta$S=1 or $\Delta$S=0). In particular, Gamow-Teller (GT) transitions ($\Delta$T=1, $\Delta$S=1, and angular momentum transfer $\Delta$L=0) have been the subject of extensive  studies. These transitions are mediated through the $\sigma\tau_{\pm}$ operator and connect the same initial and final states as $\beta_{\pm}$ decays. Weak interaction rates (via electron capture (EC) and $\beta$-decay), are pivotal in understanding the late evolution of stars \cite{fuller80,fuller82a,fuller82b,fuller85,LAN03,HEG01,heger01a,langanke03}. Since $\beta$-decay has access to states only in a very limited energy window and direct measurements with neutrinos of the full response are difficult due to the weakness of the interaction, charge-exchange reactions with hadronic probes are the preferred tool for mapping the Gamow-Teller response. 

Here, we present results for extracting Gamow-Teller strengths in the (n,p) direction using a new tool: the (t,$^{3}$He) reaction at 115 MeV/nucleon using a beam of secondary triton particles at the NSCL. In preparation for experiments on target nuclei of importance in stellar evolution (pf and sdg-shell nuclei), we chose a lighter target, $^{26}$Mg, for which the structure is well known and thus allows for a detailed study of strength-extraction techniques. The $^{26}$Mg(t,$^{3}$He)$^{26}$Na data are combined with results from a $^{26}$Mg($^{3}$He,t)$^{26}$Al experiment at 140 MeV/nucleon, performed at RCNP. Comparing transitions to analog (T=2) states excited in $^{26}$Na and $^{26}$Al provide a means for checking consistency under the assumption of isospin symmetry. In addition, $^{26}$Mg(d,$^{2}$He) \cite{XU96,NII94} and $^{26}$Mg(p,n) \cite{MAD87} data are available, allowing for a comparison with results from $^{26}$Mg(t,$^{3}$He) and $^{26}$Mg($^{3}$He,t), respectively and thus providing insight about the probe-dependence of the strength extraction.  

Detailed studies of the mechanism of the ($^{3}$He,t) and (t,$^{3}$He) reactions at energies $\sim120-140$ MeV/nucleon are rare \cite{ZEG03,FUJ04}. A good understanding is important, however. If states excited via Gamow-Teller transitions cannot be separated from states of different multi-polarity a multipole decomposition analysis (MDA) most be performed to disentangle the Gamow-Teller contribution. The MDA relies on the reliable prediction of angular distributions in the reaction models. Secondly, to estimate systematic errors in the extraction of the Gamow-Teller strengths the reaction model must incorporate the leading cause(s) for such uncertainties. Therefore, a significant portion of this paper deals with the development of treatment of the ($^{3}$He,t) and (t,$^{3}$He) reactions in the Distorted Wave Born Approximation (DWBA). The $^{26}$Mg($^{3}$He,t) data set contains many clearly separated states of varying multipolarities and thus provides an ideal testing ground for the calculations and the ability to describe the angular distributions and estimate systematic errors in the extraction of Gamow-Teller strength.   

A variety of reactions have been applied to the determination of Gamow-Teller strengths (B(GT) \cite{OST92,RAP94,HAR01}, here defined so that B(GT)=3 for the decay of a free neutron). Systematic studies of Gamow-Teller strength were first performed at IUCF (see e.g. \cite{BAI80,GOO80,GAA81,GAA85}) using the $\Delta$T$_{z}=-1$ (p,n) reaction ($T_{z}$ is the z-component of the isospin). At sufficiently high beam energies ($\sim 100$ MeV and higher), the forward-angle cross section is dominated by the $\sigma\tau$ component of the effective interaction \cite{LOV81,LOV85} that mediates Gamow-Teller transitions and a proportionality between cross sections at zero momentum transfer and Gamow-Teller strength was established \cite{TAD87}. Resolutions vary from about 200 keV for proton energies of 120 MeV to 0.65 MeV-1.9 MeV at proton energies of several hundred MeV (\cite{RAP94} and references therein, \cite{SAK96}). The (n,p) reaction was subsequently used to extract Gamow-Teller strength distributions in the inverse direction \cite{JAC88,ALF86}. The resolutions that can be obtained with (n,p) reactions are $\sim 1$ MeV. For tests of theoretically predicted Gamow-Teller strength distributions of importance for stellar evolution, data with better resolution are important since the electron-capture rates in the stellar environment are sensitive to the details of the low-lying transitions \cite{hagemann04}. Therefore, probes to extract these strengths with better resolution have been developed.

An alternative to the (n,p) reaction is the (d,$^{2}$He) reaction. Experiments have been performed at RIKEN \cite{OKA95}, Texas A\&M \cite{XU96} and at KVI \cite{RAK02} where the best resolutions have been achieved ($\sim$130 keV at E(d)=85 MeV/n). By selecting small relative energies of the outgoing protons in the unbound $^{2}$He system, enhanced selectivity for spin-transfer transitions is achieved \cite{BUG87,OKA99}. The (d,$^{2}$He) reaction has been used to extract Gamow-Teller distributions in a variety of nuclei \cite{GRE04,FRE04} focusing on cases of importance for astrophysical applications.

The ($^{3}$He,t) reaction has been used extensively to probe the spin-isospin response in nuclei. Experiments have been performed at a variety of institutions and at various beam energies (for an overview, see Ref. \cite{HAR01}). At present, the most extensive program is carried out at RCNP using the spectrometer Grand Raiden \cite{FUJ99,FUJ96}. $^{3}$He beam energies of 140-150 MeV/nucleon are used. Resolutions of 35 keV have been achieved \cite{FUJ02} and reactions on a number of targets have been performed with the main aim of extracting Gamow-Teller strength (see e.g. \cite{FUJ04a,FUJ04b,FUJ03,FUJ05}). The selectivity for Gamow-Teller transitions at forward angles and intermediate beam energies is very similar to that of the (p,n) reaction \cite{OST92}. 

The (t,$^{3}$He) reaction is a relatively new tool for studying spin-isospin excitations in the $\Delta T_{z}=+1$ reaction. The only triton beam presently available at energies above 100 MeV/nucleon is the secondary triton beam made at NSCL. The first (t,$^{3}$He) experiment, on a series of light nuclei \cite{DAI98}, was performed in a simple spectrometer and the energy resolution was poor and angular distributions could not be measured. By applying the dispersion-matching technique in the S800 spectrometer \cite{BAZ03}, it was shown that energy resolutions of $\sim200$ keV can be obtained and angular distributions be extracted \cite{SHE99}. Subsequently, states in weakly-bound $^{6}$He via $^{6}$Li(t,$^{3}$He) \cite{NAK00} were studied. 

This paper is structured as follows: In Section \ref{sec:prel} the framework for extracting Gamow-Teller strengths from the ($^{3}$He,t) and  (t,$^{3}$He) reactions and the motivation for choosing $^{26}$Mg as a target will be will be discussed. In Section  \ref{sec:experiment}, the details of the $^{26}$Mg($^{3}$He,t) and $^{26}$Mg(t,$^{3}$He) experiments and the general features of the measured spectra are described. In Section \ref{sec:strength}, the strengths extracted from the data are presented and compared with results from other charge-exchange experiments and with shell-model calculations. In addition, the exhaustion of the model-independent Gamow-Teller sum rule is discussed. Since the reaction calculations in DWBA that are employed in this section are novel for ($^{3}$He,t) and  (t,$^{3}$He) reactions, they are discussed in detail in Section \ref{sec:dwba} and subsequently employed to describe the systematic uncertainties in the extraction of Gamow-Teller strength from the data. 

\begin{figure}
\includegraphics[width=8.5cm]{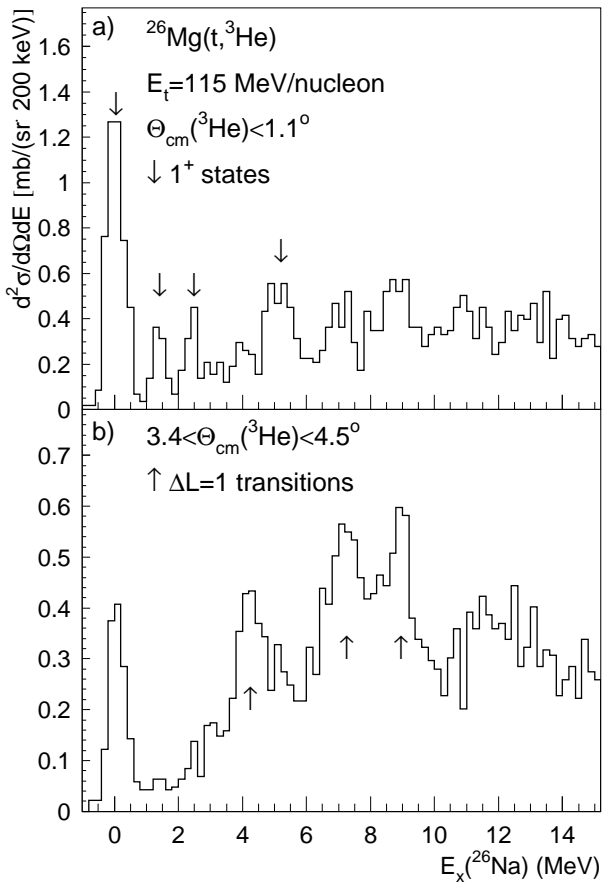}
\caption{\label{prc1} Differential cross sections measured for the $^{26}$Mg(t,$^{3}$He) reaction at E$_{t}$=115 MeV/nucleon. a) The spectrum for $\theta_{cm}(^{3}\text{He})<1.1^{\circ}$. b) The spectrum for $3.4^{\circ}<\theta_{cm}(^{3}\text{He})<4.5^{\circ}$. Unlike transitions with higher angular momentum transfer, Gamow-Teller ($\Delta$L=0) transitions peak at forward angles. The energy region where Gamow-Teller transitions dominate the spectrum is indicated in a), and the transitions to the $1^{+}$ states are indicated by down arrows. In contrast, dipole transitions ($\Delta$L=1) have a minimum at forward angles and peak near 3.5$^{\circ}$. Such transitions are indicated by up arrows in b). 
}
\end{figure}

\begin{figure}
\includegraphics[width=8.5cm]{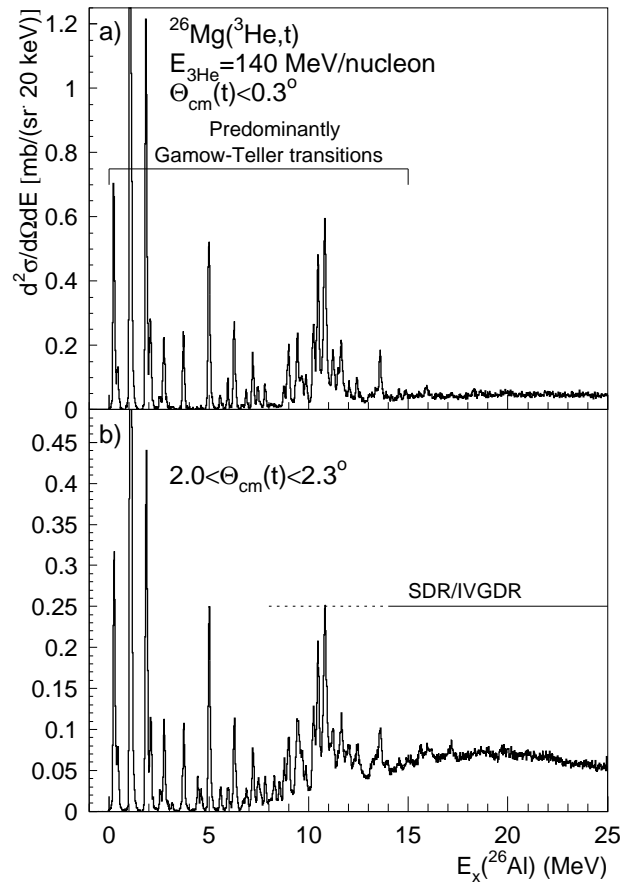}
\caption{\label{prc2} Differential cross sections measured for the $^{26}$Mg($^{3}$He,t) reaction at E$_{^{3}\text{He}}$=140 MeV/nucleon. The top panel shows the spectrum for $\theta_{cm}(t)<0.3^{\circ}$. The bottom panel shows the spectrum for $2.0^{\circ}<\theta_{cm}(t)<2.3^{\circ}$. As in Fig. 1, transitions with $\Delta$L=0 peak at forward angles. The broad bump in the spectrum, seen at larger angles (indicated in the bottom panel) is due to giant dipole resonances (IVSDR and IVGDR).}
\end{figure}

\section{Preliminaries}
\label{sec:prel}
In this article we present measurements of the $^{26}$Mg($^{3}$He,t) and $^{26}$Mg(t,$^{3}$He) reactions at 140 MeV/nucleon and 115 MeV/nucleon, respectively. The $^{26}$Mg nucleus provides an excellent case for testing the reaction mechanism. Its structure is well known, and full sd-shell wave functions are available, a prerequisite for the studies of the reaction mechanism. Here, we test the new USD interaction, USD-05B \cite{USDN}. The isospin of $^{26}$Mg is T=1. Thus, final states with T=0,1 and 2 transitions can be observed in the $^{26}$Mg($^{3}$He,t) reaction. The results for the T=2 states can be compared with their analog states excited via (t,$^{3}$He) to verify consistency. Assuming isospin symmetry, the strengths of analog transitions differ only by their Clebsch-Gordan coefficients; in this case, $\frac{\text{B(GT)}_{^{26}\text{Mg}}(^{3}\text{He},t)}{\text{B(GT)}_{^{26}\text{Mg}}(t,^{3}\text{He})}=\frac{1}{6}$. Note that a comparison between analog Gamow-Teller transitions becomes more difficult for target nuclei with increasing ground state isospin (larger N-Z) since the ratio $\frac{\text{B(GT)}(\Delta T_{z}=-1)}{\text{B(GT)}(\Delta T_{z}=+1)}$ decreases, and the transitions in the $\Delta T_{z}=-1$ direction become more difficult to separate from the continuum and other transitions.   

The $^{26}$Mg($^{3}$He,t) experiment is also important for checking that the extracted Gamow-Teller strengths are consistent with those from (p,n) and, therefore, probe independent.
Since $^{26}$Al has isospin T=0, the Gamow-Teller strengths for the transitions to the four lowest-lying 

$1^{+}$ states can be extracted from $^{26}$Si $\beta^{+}$-decay \cite{WIL80} data if the small contribution from isospin symmetry breaking is neglected. It can then be used to calibrate the proportionality between Gamow-Teller strength and cross sections at zero momentum transfer (q=0).
In eikonal approximation this proportionality is written as \cite{TAD87}:
\begin{equation}
\label{eq:eik}
\frac{d\sigma}{d\Omega}(q=0)=KN|J_{\sigma\tau}|^{2}B(GT)=\hat{\sigma}B(GT).
\end{equation} 
Here, $K=\frac{E_{i}E_{f}}{(\hbar^{2}c^{2}\pi)^{2}}$ where $E_{i(f)}$ is the reduced energy in the incoming (outgoing) channel; $N$ is the distortion factor defined by the ratio of distorted-wave to the plane-wave cross sections;  $|J_{\sigma\tau}|$ is the volume-integral of the central $\sigma\tau$ interaction. The factor $KN|J_{\sigma\tau}|^{2}$ is referred to as the unit cross section, $\hat{\sigma}$. 
The cross section for momentum transfer q=0, requiring both the Q-value of the transition and the scattering angle to be zero, is obtained by extrapolating the data using:
\begin{equation}
\label{eq:extra}
\frac{d\sigma}{d\Omega}(q=0)=\left[\frac{\frac{d\sigma}{d\Omega}(q=0)}{\frac{d\sigma}{d\Omega}(Q,0^{\circ})}\right]_{\text{DWBA}}\times\left[\frac{d\sigma}{d\Omega}(Q,0^{\circ})\right]_{\text{exp}}.
\end{equation}
In this equation, `$\text{DWBA}$' refers to calculated values in the Distorted-Wave Born Approximation described in section \ref{sec:dwba}. The experimental cross section at $\theta=0^{\circ}$ is obtained by fitting the calculated Gamow-Teller angular distribution in DWBA to the measured angular distribution. The proportionality of Eq. \ref{eq:eik} also holds for Fermi transitions \cite{TAD87} associated with the excitation of the  
isobaric analog state (IAS) via $\Delta T_{z}=-1$ reactions. In that case,  $|J_{\sigma\tau}|$ has to be replaced by $|J_{\tau}|$.

Besides the trivial dependence of $K$ on target and projectile mass and beam energy, also N and $J_{\sigma\tau}$ depend on mass and beam energy. The distortion factor $N$ can be calculated if the optical potential parameters are known. In general, the distortion factors for the ($^{3}$He,t) and (t,$^{3}$He) reactions are smaller (meaning larger distortions) than those for (p,n) and (n,p), respectively, since the latter two probe the nuclear interior more strongly than the reactions with composite probes. The $\sigma\tau$ component of the interaction ($J_{\sigma\tau}$) is only weakly dependent on the beam energy \cite{LOV81,LOV85}, but due to the exchange terms, the dependence on target mass is significant, especially for lower mass numbers. 

In practice, the proportionality described by Eq. \ref{eq:eik} is often calibrated by correlating extracted cross sections at zero momentum transfer with empirically known Gamow-Teller strengths from $\beta$-decay data.  

In the Love-Franey effective interaction \cite{LOV81,LOV85}, which we will employ in DWBA cross section calculations for more detailed comparisons between data and theory, the central $\sigma\tau$ and $\tau$ interactions are represented by superpositions of real and imaginary Yukawa potentials with varying ranges corresponding to the different meson-exchange potentials. Spin-orbit contributions are represented by a summation over Yukawa potentials multiplied by the $\mathbf{L \cdot S}$ operator. The main source of systematic errors in the proportionality of Eq. \ref{eq:eik} is interference from $\Delta$L=2, $\Delta$S=1 amplitudes for transitions to the $J^{\pi}=1^{+}$ Gamow-Teller states, mediated mainly through the tensor-$\tau$ component of the interaction. In the Love-Franey effective interaction, this component is represented by a summation over potentials of the form $r^{2}\times\text{Yukawa}$ multiplied by the $S_{12}=\frac{(\mathbf{\sigma_{1}\cdot r})(\mathbf{\sigma_{2}\cdot r})}{r^{2}}-\mathbf{\sigma_{1}\cdot\sigma_{2}}$ operator. The cross-section calculations and such systematic errors will be discussed further in section \ref{sec:dwba}. At beam energies above $\sim100$ MeV/nucleon the contributions from multistep processes become small \cite{FUJ96} and are less of a concern in terms of the breaking of proportionality and are disregarded in this work.

\section{Experiments}
\label{sec:experiment}

\subsection{The $^{26}$Mg(t,$^{3}$He) experiment at NSCL}
\begin{figure*}
\includegraphics[scale=0.85]{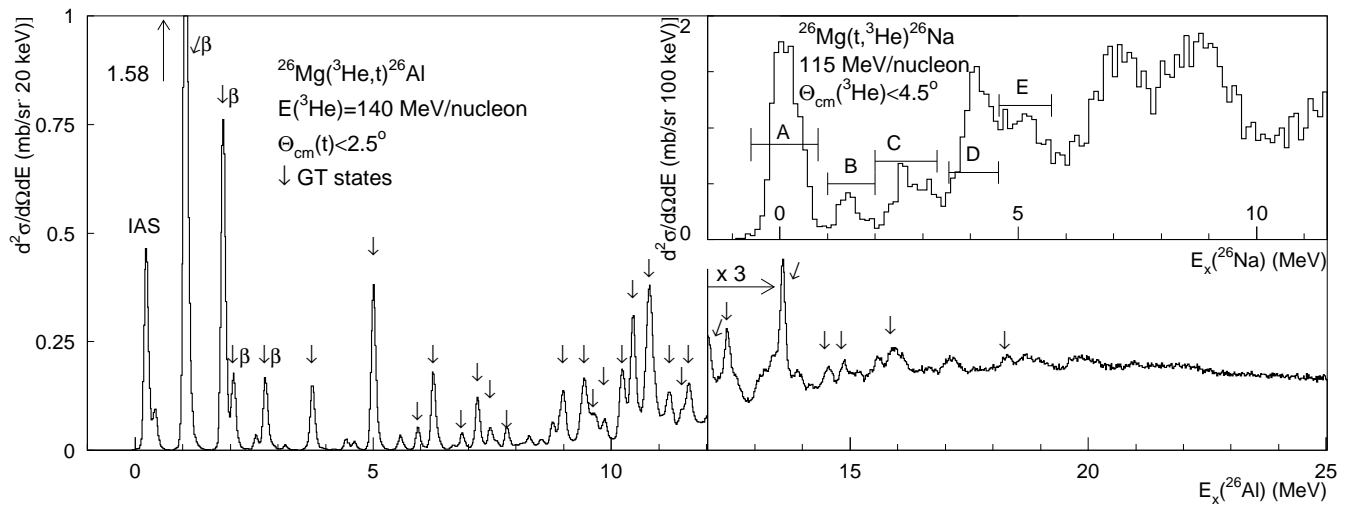}
\caption[]{\label{prc3}Energy spectra of the $^{26}$Mg($^{3}$He,t) and $^{26}$Mg(t,$^{3}$He) reactions. Peaks identified as due to Gamow-Teller transitions are indicated. `$\beta$' denotes that the Gamow-Teller strength of the transition can be deduced from $^{26}$Si $\beta$-decay under the assumption of isospin symmetry. The $^{26}$Mg(t,$^{3}$He) spectrum has been shifted by the Coulomb-energy difference so that the 1$^{+}$ state at $E_{x}$=0.08 keV is aligned with its analog at E$_{x}$=13.6 MeV in the $^{26}$Mg($^{3}$He,t) spectrum. }
\end{figure*}

The production of the secondary triton beams for (t,$^{3}$He) experiments is described in detail in Refs. \cite{DAI97,SHE99} and here we only give an overview and refer to circumstances specific for taking the $^{26}$Mg(t,$^{3}$He) data.  A secondary triton beam with an average energy of 115-MeV/nucleon and a relative energy spread of 1\% was produced from a primary 140 MeV/nucleon $\alpha$-beam impinging on a 9.25- g/cm$^{2}$ thick Be production target. The primary beam was accelerated in the K1200 cyclotron and the tritons were selected in the A1200 fragment separator. The triton intensity ($~1.3\times 10^{6}$ tritons/s) was monitored with an in-beam scintillator (IBS) placed 30 m upstream from the isotopically-enriched (99.42\%) $12.5\pm0.3$-mg/cm$^{2}$ thick $^{26}$Mg target. The $^{3}$He ejectiles were detected in the S800 spectrometer \cite{BAZ03}, set at an angle of 0$^{\circ}$ and operated in dispersion-matched mode. Since the triton beam cannot be bent into the focal plane of the spectrometer, the transmission of the triton beam from the IBS to the target was determined by comparing the rates at the S800 focal plane and the IBS using a secondary $^{3}$He beam without a target. Even though the beam-spot size in the dispersive direction is relatively large ($\sim 5$cm), a transmission of 95\% was achieved owing to the large acceptance of the S800 spectrometer. 

The S800 focal-plane detector system \cite{YUR99} consists of two 2-dimensional cathode-readout drift detectors (CRDCs) for the determination of position and angles in the focal plane. In addition, two thin plastic scintillators are positioned behind the CRDCs.  The first one serves as trigger for the data acquisition system and provides the start signal for the time-of-flight (TOF) measurement. The stop signal is provided by the cyclotron radio frequency. The light output of the two scintillators and the TOF signal were used to identify $^{3}$He particles. For ray-tracing purposes, the ion-optical code COSY Infinity \cite{COSY} was used to calculate the ion-optical transfer matrix of the S800  spectrometer \cite{BER93} from the measured magnetic field maps. Matrix elements up to fifth order were used in the reconstruction of $\delta=(E-E_{0})/E_{0}$ ($E_{0}$ is the kinetic energy of the particle following the central-ray trajectory through the spectrometer and $E$ the energy of the measured particle). The track angles in the dispersive and non-dispersive directions were also obtained in the ray-tracing procedure and used to calculate the $^3$He scattering angle.
The $^{26}$Na excitation energy was determined in a missing-mass calculation using the reconstructed energy and scattering angle. The overall energy resolution was 300 keV, full-width at half maximum (FWHM). This is about 100 keV worse than the resolution achieved in the experiment described in Ref. \cite{SHE99}, due to the relatively thick target used in this experiment and slightly poorer dispersion matching conditions. Cross sections were measured at center-of-mass angles from $0^{\circ}$ to $4.5^{\circ}$. The angular resolution was 0.4$^{\circ}$ (FWHM). 

\subsection{The $^{26}$Mg($^{3}$He,t) experiment at RCNP}
The $^{26}$Mg($^{3}$He,t) data were obtained at RCNP using the spectrometer Grand Raiden \cite{FUJ99}. A 3-pnA 140-MeV/nucleon $^{3}$He beam bombarded a 3.6 mg/cm$^{2}$ isotopically enriched (99.4\%) $^{26}$Mg target. The spectrometer was run in the ``off-focus'' mode \cite{FUJ01} so that the angle in both horizontal and vertical directions could be determined. The beam current was integrated using a Faraday cup. The experimental setup and analysis techniques were the same as described in Refs. \cite{ZEG03,ZEG04a}. 
The energy resolution was $\sim$100 keV (FWHM). As discussed in the introduction, Gamow-Teller strengths have been extracted from high-resolution ($\sim30$ keV) experiments at RCNP using the dispersion-matching technique on a variety of targets, including $^{26}$Mg \cite{FUJ03} (up to $E_{x}$($^{26}$Al)=9 MeV). In the high-resolution measurements, absolute cross sections and precise angular distributions were not extracted.  Owing to the increased beam-spot size, the acceptance of the spectrometer (which is much smaller than that of the S800 spectrometer) is not defined by the entrance slits in a simple way and accurate beam current integration is problematic. Therefore, for experiments performed in dispersion-matched mode, the yield in a narrow angular bin (typically from $0^{\circ}$ to $\sim 0.5^{\circ}$, see e.g. Ref. \cite{FUJ03}) is used in Eq. \ref{eq:extra} and the identification of Gamow-Teller transitions performed by comparing ratios of yields at forward and backward angles.
For the purpose of studying the details of the reaction mechanism, the absolute cross sections and accurate angular distributions are essential. The ($^{3}$He,t) data presented here were, therefore, taken in achromatic mode where beam integration is straightforward and the acceptance is well defined due to the small beam-spot size on the target of about 1-2 mm. Center-of-mass scattering angles were measured from $0^{\circ}$ to $2.5^{\circ}$, with a resolution of $0.2^{\circ}$ (FWHM).   

As a check on the reliability of the beam integration for the ($^{3}$He,t) experiment presented here, some data were also taken with $^{13}$C and $^{208}$Pb targets under exactly the same conditions as for the $^{26}$Mg target. The differential cross sections for the transitions to the ground state of $^{13}$N \cite{FUJ04} and the isobaric analog state (IAS) in $^{208}$Bi \cite{ZEG03} for these two targets have been extracted in the past. The results from the previous and new data differed by less than 10\%, giving a measure for the systematic uncertainties in the beam integration and target thickness.   

\subsection{The measured spectra and their general features}
In Figs. \ref{prc1} and \ref{prc2}, the energy spectra measured via $^{26}$Mg(t,$^{3}$He) and $^{26}$Mg($^{3}$He,t), respectively, are shown. Transitions with $\Delta L=0$ peak at $0^{\circ}$ and drop off rapidly. They dominate the spectra at forward angles; see Figs. \ref{prc1}a and \ref{prc2}a in which states excited via monopole transitions are marked by a down arrow. In contrast, dipole transitions peak at around 3$^{\circ}$ and are thus enhanced at backward angles; see the Figs. \ref{prc1}b and \ref{prc2}b where states excited via dipole transitions are marked by an up arrow. 
Transitions with higher angular-momentum transfer are possible as well, but they have a rather flat angular distribution over the angular range measured here.

In the $\Delta T_{z}=+1$ direction, only the $\Delta L=0$ isovector (spin-flip) giant monopole resonances have angular distributions similar to the Gamow-Teller transitions. These $2\hbar\omega$ resonances have high excitation energies ($\gtrsim20$ MeV \cite{AUE83}). For excitation energies below $\sim20$ MeV, identification of Gamow-Teller transitions using the angular distributions is very selective. 
Therefore, the states in Fig. \ref{prc1}a indicated with a down arrow are identified as Gamow-Teller transitions to 1$^{+}$ states in $^{26}$Na. The detailed procedure of extracting components in the spectrum using a MDA is discussed in Section \ref{sec:strength}. 

The dipole transitions visible in Fig. \ref{prc1}b, constitute the various components ($J^{\pi}=0^{-}, 1^{-}, 2^{-}$) of the isovector spin-flip giant resonance (SDR; $\Delta L=1$, $\Delta S=1$) and its non-spin-transfer partner, the $J^{\pi}=1^{-}$ isovector giant dipole resonance (IVGDR; $\Delta L=1$, $\Delta S=0$). Relative to the IVSDR, the IVGDR is weak, because non-spin-transfer transitions are suppressed at these beam energies \cite{LOV81,LOV85}. 

For the $\Delta T_{z}=-1$ direction, the angular distributions can similarly be used for unambiguous identification of Gamow-Teller transitions. The isovector giant monopole resonances are located at energies above 30 MeV \cite{AUE83}). There is one additional transition with $\Delta L=0$ in this direction, namely the excitation of the IAS ($\Delta L=0, \Delta S=0$). Its location is well known, however, and its strength can be removed easily in the extraction of Gamow-Teller strengths.
In the $^{26}$Mg($^{3}$He,t) data (Fig. \ref{prc2}) many Gamow-Teller transitions can easily be identified. Since only center-of-mass angles of up to $2.5^{\circ}$ are measured, even at the most backward angles (Fig. \ref{prc2}b), they dominate the spectrum. The dipole contributions (due to the IVSDR and IVGDR) appear mostly as a broad resonance with a maximum at an energy of about 20 MeV. The good resolution of 100 keV allows for a peak-by-peak analysis, instead of a MDA. The peak shape was determined using the strongest Gamow-Teller transition at $E_{x}$=1.06 MeV and was then used in the yield estimation for all transitions. If a peak was not isolated, the background under it was parameterized with a polynomial in the energy region close to the peak and a systematic error to the yield was assigned based on the ambiguity in estimating the background. If two or more peaks were not completely separated, the fits were performed simultaneously for the peaks in that region. The data set was divided into eight angular regions and the yields for all peaks were obtained in each region separately.   

Note that at our beam energies and at forward angles, transitions that involve large angular-momentum transfers are strongly suppressed. A good example is the transition to the $5^{+}$ ground state of $^{26}$Al, for which no statistically significant yield was measured whereas at lower $^{3}$He beam energies (e.g. 24 MeV/nucleon \cite{GRA86}) this state can be observed clearly. 

\section{Extraction of Gamow-Teller strengths and comparison with shell-model calculations}
\label{sec:strength}
\begin{figure}
\includegraphics[width=8.5cm]{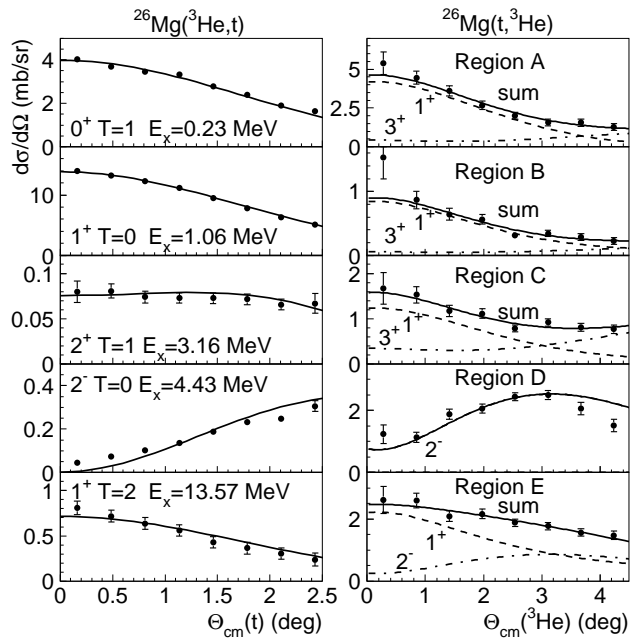}
\caption{\label{prc4}Measured differential cross sections and comparison with DWBA calculations. Left: Results for 5 states in the $^{26}$Mg($^{3}$He,t) spectrum with different J$^{\pi}$ and T are shown. Right: Results of MDA in regions A-E of Fig. \ref{prc3} in the $^{26}$Mg(t,$^{3}$He) spectrum are shown.
}
\end{figure}
Fig. \ref{prc3} shows the measured energy spectra integrated over the full angular ranges covered in the two experiments. The $^{26}$Mg(t,$^{3}$He) spectrum has been shifted by the Coulomb-energy difference so that the first Gamow-Teller state at $E_{x}=0.08$ keV is aligned with the first T=2 Gamow-Teller state at E$_{x}$=13.6 MeV in the $^{26}$Mg($^{3}$He,t) spectrum. 

\subsection{Gamow-Teller strength in $^{26}$Al}
For the $^{26}$Mg($^{3}$He,t) data, Gamow-Teller states were identified using their $L=0$ nature by comparing measured angular distributions for each peak in the spectrum to DWBA calculations performed with the code FOLD \cite{FOLD} (see section \ref{sec:dwba}). Fig. \ref{prc4} shows a comparison of the measured ($^{3}$He,t) differential cross sections and DWBA calculations of transitions to several states of various multipolarity. For each transition, the only freedom in comparing the data with the DWBA calculations was an angle-independent scale factor, which was determined in a fit. The comparison is shown for the $0^{+}$ IAS at $E_{x}$=0.23 MeV, the strongest $1^{+}$ state at $E_{x}$=1.06 MeV, a $2^{+}$ state at $E_{x}$=3.16 MeV, a $2^{-}$ state at $E_{x}$=4.43 MeV and the first T=2 $1^{+}$ state at $E_{x}$=13.57 MeV. In all cases a good correspondence between the data and calculation is found, giving confidence that angular distributions can be predicted accurately using the model. For the states in the spectrum that are identified as 1$^{+}$ states, the zero-degree cross section was extracted from the fitted theoretical curve, with the error being deduced from the fitting error. Eq. \ref{eq:extra} was then used to obtain the cross section at zero-momentum transfer. In total, 29 Gamow-Teller transitions were identified. In Fig. \ref{prc3}, they are marked with arrows. Under the assumption of isospin symmetry, the Gamow-Teller strengths for the first four of these Gamow-Teller transitions can be deduced from $\beta$-decay measurements of $^{26}$Si \cite{WIL80}.  The corresponding B(GT)s are 1.098, 0.536, 0.091 and 0.113 for E$_{x}$($^{26}$Al)=1.06 MeV, 1.85 MeV, 2.07 MeV and 2.74 MeV, respectively. For reasons discussed in the next section, only the first two (with the larger values of Gamow-Teller strength) of these transitions were used to determine the unit cross section $\hat{\sigma}$ in Eq. \ref{eq:eik}. This unit cross section was then used to calculate the Gamow-Teller strengths for all 29 peaks. The results are summarized in Table \ref{tab:table1}, combined with previous results from the high-resolution $^{26}$Mg($^{3}$He,t) experiment \cite{FUJ03}, the results from a $^{26}$Mg(p,n) experiment \cite{MAD87} at $E_{p}=135$ MeV with an energy resolution of 300 keV and the strengths extracted from $\beta$-decay, following Ref. \cite{FUJ03}. Note that in the analysis of the high-resolution ($^{3}$He,t) data and the (p,n) data, all four Gamow-Teller transitions for which the strength is known from $\beta$-decay, were used in the calibration of the unit cross section of Eq. \ref{eq:eik}. 

\begin{table*}
\caption{\label{tab:table1} Extracted Gamow-Teller strengths for transitions from $^{26}$Mg to $^{26}$Al from the present $^{26}$Mg($^{3}$He,t) experiment at 140 MeV/nucleon, a high-resolution $^{26}$Mg($^{3}$He,t) experiment at 140 MeV/nucleon \cite{FUJ03}, a $^{26}$Mg(p,n) experiment at 135 MeV/nucleon \cite{MAD87} and from $\beta$-decay measurements of $^{26}$Si \cite{WIL80} assuming isospin symmetry. The error bars for the present ($^{3}$He,t) data set given in the table include statistical errors, as well as systematical errors in cases where background had to be subtracted. Estimates of other systematic errors are given in section \ref{sec:dwba}.}
\begin{ruledtabular}
\begin{tabular}{cccccccc}
\multicolumn{2}{c}{$\beta$-decay\footnotemark[1]} &\multicolumn{2}{c}{($^{3}$He,t) present data} &\multicolumn{2}{c}{($^{3}$He,t) high resolution \footnotemark[2]} &\multicolumn{2}{c}{(p,n)\footnotemark[3]} \\ \hline
E$_{x}$ (MeV)& B(GT)           & E$_{x}$ (MeV) & B(GT)          & E$_{x}$ (MeV) & B(GT)          & E$_{x}$ (MeV) & B(GT) \\ \hline
1.0577       & 1.098$\pm$0.022 & 1.06          &1.09$\pm$0.03\footnotemark[5]   &1.058          &1.081$\pm$0.029\footnotemark[5] &1.06           & 1.10\footnotemark[5]\\
1.8506       & 0.536$\pm$0.014 & 1.85          &0.54$\pm$0.02\footnotemark[5]   &1.850          &0.527$\pm$0.015\footnotemark[5] &1.85           & 0.50\footnotemark[5]\\
2.0716       & 0.091$\pm$0.004 & 2.07          &0.114$\pm$0.008 &2.071          &0.112$\pm$0.004\footnotemark[5] &2.11           & 0.11\footnotemark[5]\\
2.7400       & 0.113$\pm$0.005 & 2.74          &0.119$\pm$0.008 &2.739          &0.117$\pm$0.004\footnotemark[5] &2.72           & 0.13\footnotemark[5]\\
             &                 & 3.73          &0.109$\pm$0.008 &3.726          &0.106$\pm$0.004 &3.73           & 0.10\\
             &                 & 5.01          &0.28$\pm$0.01   &5.010          &0.271$\pm$0.008 &5.01           & 0.28\\
             &                 & 5.94          &0.041$\pm$0.005 &5.949          &0.037$\pm$0.002 &(5.95)         & 0.04\\
             &                 & 6.27          &0.134$\pm$0.008 &6.269          &0.126$\pm$0.004 &(6.28)         & 0.12\\
             &                 & 6.87          &0.028$\pm$0.004 &6.875          &0.028$\pm$0.001 &(6.87)         & 0.03\\
             &                 & 7.20          &0.089$\pm$0.006 &7.199          &0.085$\pm$0.003 &(7.21)         & 0.09\\
             &                 & 7.46          &0.036$\pm$0.004 &7.457          &0.038$\pm$0.002 &               & \\
             &                 & 7.81          &0.037$\pm$0.004 &7.813          &0.040$\pm$0.002 &(7.85)         & 0.04\\
             &                 & 8.98          &0.123$\pm$0.008 &8.930          &0.041$\pm$0.002\footnotemark[4] &(8.94)         &0.15 \\
             &                 &               &                &9.007          &0.079$\pm$0.003\footnotemark[4] &               & \\ \hline
\multicolumn{2}{c}{Sum for E$_{x}<$9.2 MeV}    &   &2.73$\pm$0.04               &   &2.69$\pm$0.04               & & 2.69\\ \hline             
             &                 & 9.43          &0.136$\pm$0.008 &               &                &               & \\
             &                 & 9.62          &0.079$\pm$0.007 &               &                &(9.77)         & 0.09\\
             &                 & 9.86          &0.058$\pm$0.006 &               &                &               & \\
             &                 & 10.24         &0.158$\pm$0.009 &               &                &(10.2)         & 0.16\\
             &                 & 10.45         &0.29$\pm$0.01   &               &                &               & \\
             &                 & 10.81         &0.47$\pm$0.02   &               &                &(10.8)         & 0.44\\
             &                 & 11.22         &0.164$\pm$0.009 &               &                &(11.2)         & 0.17\\
             &                 & 11.50         &0.021$\pm$0.005 &               &                &               & \\
             &                 & 11.62         &0.17$\pm$0.01   &               &                &(11.6)         & 0.20\\
             &                 & 12.01         &0.015$\pm$0.003 &               &                &               & \\
             &                 & 12.41         &0.022$\pm$0.004 &               &                &(13.1)         & 0.05\\
             &                 & 13.57         &0.068$\pm$0.003 &               &                &(13.6)         & 0.12\\
             &                 & 14.53         &0.015$\pm$0.004 &               &                &(14.6)         & 0.11\\
             &                 & 14.88         &0.018$\pm$0.005 &               &                &(14.9)         & 0.07\\
             &                 & 15.91         &0.029$\pm$0.006 &               &                &               & \\
             &                 & 18.32         &0.021$\pm$0.005 &               &                &               & \\ \hline
Sum          &                 &               &  4.46$\pm$0.05 &               &                &               & 4.41\\

\end{tabular}
\end{ruledtabular}
\footnotetext[1]{From Ref. \cite{WIL80}, assuming isospin symmetry}
\footnotetext[2]{From Ref. \cite{FUJ03}}
\footnotetext[3]{From Ref. \cite{MAD87}; we follow the notation by the authors to put parentheses around excitation energy values that may represent an average over more than one state.}
\footnotetext[4]{These two states are unresolved in the present ($^{3}$He,t) data set. The sum of the strengths (0.120$\pm$0.004) is consistent with the value of 0.123$\pm$0.008 extracted in the present data.}
\footnotetext[5]{Used in the calibration of the unit cross section in Eq. \ref{eq:eik}.}
\end{table*}

The errors in table \ref{tab:table1} of the present ($^{3}$He,t) data include a statistical as well as a systematical component resulting from uncertainties in the background estimates in cases where peaks were not separated. Systematical errors that cannot be estimated from the data are discussed in section \ref{sec:dwba}.
A state at $E_{x}=5.58$ MeV, was seen in both the present ($^{3}$He,t) and the high-resolution data. It was listed in the compilation of Ref. \cite{END90} as a $1^{+}$ state, but in the updated compilation of Ref. \cite{END98} no parity assignment was shown. In Ref. \cite{FUJ03} the transition to this state was assigned to have $\Delta L \neq 0$, but nevertheless given a tentative Gamow-Teller strength of 0.020$\pm$0.001. We found that the angular distribution did not peak strongly at $0^{\circ}$, as expected for Gamow-Teller transitions. However, it cannot be ruled out that two unresolved states with different multipolarities make up this peak. Since the strength of this transition is small compared to the sum of the Gamow-Teller strengths of all transitions as well as its error, it was excluded from Table \ref{tab:table1}.     

Overall, a good correspondence is found between the extracted Gamow-Teller strengths below $E_{x}$=9.2 MeV from the three charge-exchange experiments compared in Table \ref{tab:table1}; The integrated sums are consistent within error margins, showing that the strength extraction is largely probe-independent. Above $E_{x}=9.2$ MeV, and in particular above $E_{x}=12$ MeV, larger differences are seen between the ($^{3}$He,t) results (present data set only) and the (p,n) results. In this energy range, the improved resolution becomes increasingly important, since it reduces the systematic errors due to the background estimates and subtraction. In Ref. \cite{MAD87}, systematic errors in the background subtraction for the (p,n) data in the energy region above $E_{x}=13.3$ MeV were estimated to be as much as 50\%, depending on the method used. The tabulated Gamow-Teller strengths of Ref. \cite{MAD87} were the ones extracted with the lowest background estimates and, therefore, highest cross section and strength. 

\subsection{Gamow-Teller strength in $^{26}$Na}
The $^{26}$Mg(t,$^{3}$He) reaction data do not allow a peak-by-peak analysis because transitions to states with different $J^{\pi}$ are not clearly resolved. An MDA is required. Since the DWBA calculations for experimentally well-separated states accurately predict angular distributions for transitions of various spin and angular-momentum transfers for the ($^{3}$He,t) reaction, we used the DWBA calculations to predict the various components used in the MDA fitting procedure. The $^{26}$Na spectrum up to $E_{x}$=6 MeV was divided into 5 regions (A-E, as shown in the inset in Fig. \ref{prc3}a). Regions A, B, C, and E were chosen where the Gamow-Teller transitions were seen to be strong based on the comparison of spectra at different angles in Fig. \ref{prc1}. Additionally, the region around the lowest-lying dipole state (region D) was analyzed. Above $E_{x}$=6 MeV, no statistically significant Gamow-Teller strength was found.

In the right-hand panel of Fig. \ref{prc4}, angular distributions extracted from the (t,$^{3}$He) data are shown for each of the regions A to E.  Several $2^{+},3^{+},4^{+},5^{+}$ states that cannot be resolved are known to reside in the region below $E_{x}$=3.5 MeV \cite{PEA87}. Their angular distributions near $0^{\circ}$ are flat and similar. At the beam energy of 115 MeV/nucleon, transitions involving highest angular-momentum transfers ($4^{+},5^{+}$) are suppressed.  In regions A-C, the MDA results shown in Fig. \ref{prc4} included transitions to $1^{+}$ and $3^{+}$ states. Separate MDAs were performed in which the transitions to $2^{+}$ states were used instead of to $3^{+}$ states. No significant difference in the extraction of Gamow-Teller strength was noted. As shown in  Fig. \ref{prc4}, the angular distribution in region D can be reproduced by assuming only the presence of dipole transitions. Therefore, in Region E a mixture of dipole and Gamow-Teller strength was assumed which led to a good fit to the data.   

The cross sections at 0$^{\circ}$ for the Gamow-Teller components found in the MDA were inserted into Eq. \ref{eq:extra} to calculate the cross section at zero-momentum transfer. Under the assumption of isospin symmetry, the cross section for the transition to the first $1^{+}$ state in $^{26}$Na is a factor of 6 larger than the cross section for the excitation of its analog T=2 state excited via ($^{3}$He,t) located at 13.57 MeV. Due to the 25 MeV/nucleon difference in beam energy between the two experiments, a kinematical correction factor ($k$) is required when comparing results. This factor was determined in DWBA to be 1.2 (to be multiplied with the cross section found via the (t,$^{3}$He) reaction). The ratio $\frac{k\sigma_{(\text{t},^{3}\text{He})}}{\sigma_{(^{3}\text{He},\text{t})}}(q=0)$ was $6.1\pm0.5$, which is consistent with the expected factor of 6 due to the difference in B(GT) arising from the isospin Clebsch-Gordan coefficients. Therefore, except for the factor $k$, the proportionality factor between Gamow-Teller strength and cross section at zero-momentum transfer used for the (t,$^{3}$He) reaction was the same as that for the ($^{3}$He,t) reaction. 

In Table \ref{tab:table2}, the extracted Gamow-Teller strengths are given and compared with results from two (d,$^{2}$He) experiments \cite{NII94,XU96}. For each of the $\Delta T_{z}=+1$ reactions, the energy of the first state is fixed to the known energy of the first $1^{+}$ state in $^{26}$Na at $E_{x}$=0.08 MeV \cite{END98}. The ground state of $^{26}$Na has spin-parity $J^{\pi}$=$3^{+}$. A $1^{+}$ state is also reported at $E_{x}$=0.23 MeV \cite{END98} from $\beta$-decay of $^{26}$Ne (B(GT)=0.04 relative to B(GT)=0.49 for the $\beta$-decay to the state at $E_{x}$=0.08 MeV). In $^{26}$Mg(t,$^{3}$He) data taken at 12 MeV/nucleon \cite{PEA87}, a state was found at this energy, and assigned $2^{+}$ based on the angular distribution. Although we cannot separate this state in the current data, a strong transition to this state would have resulted in a significant broadening of the first peak in the $^{26}$Na energy spectrum. Also, no transition to the T=2 analog in $^{26}$Al via ($^{3}$He,t) was obeserved, although it would certainly be resolvable. Therefore, we conclude that, if the state at $E_{x}$=0.23 MeV is excited at all, the B(GT) for this transition must be small. We assign error bars to the energies of the $1^{+}$ states at $E_{x}$=1.4 MeV, 2.6 MeV and 5.1 MeV extracted from the (t,$^{3}$He) experiment, based on the energy uncertainties of the relevant peaks in the spectrum at forward angles (Fig. \ref{prc1}a).     

The first of the two (d,$^{2}$He) experiments \cite{NII94} was performed at 135 MeV/nucleon with an energy resolution of 650 keV. The Gamow-Teller distribution extracted is similar to our result using the (t,$^{3}$He) reaction, but when summed, 25$\pm$2\% less strength is recovered. No explicit error margins are given in Ref. \cite{NII94}, so the error in this discrepancy is calculated from the (t,$^{3}$He) result only. The authors assign a general uncertainty in the extracted Gamow-Teller strengths based on the analysis of a number of nuclei of about 20\%. In the second (d,$^{2}$He) measurement \cite{XU96}, performed at 63 MeV/nucleon with an energy resolution of 650 keV, the Gamow-Teller strength for the first $1^{+}$ state is consistent with our result from (t,$^{3}$He). 
It should be noted that in neither (d,$^{2}$He) experiment a MDA analysis was performed. $^{26}$Mg(n,p) data also exist \cite{YEN87}, but the resolution was relatively poor and strengths were not extracted. The qualitative features of those data are similar to the (t,$^{3}$He) and (d,$^{2}$He) results, however. 

In Table \ref{tab:table2}, candidate T=2 Gamow-Teller states from the $^{26}$Mg($^{3}$He,t) are also included. Except for one state at $E_{x}$=14.53 MeV in $^{26}$Al, presumably a high-lying T=1, $1^{+}$ state, a good one-to-one correspondence is found, both in terms of energies of the states and their strengths. 

\begin{table*}
\caption{\label{tab:table2} Extracted Gamow-Teller strengths for transitions from $^{26}$Mg to $^{26}$Na from the present $^{26}$Mg(t,$^{3}$He) experiment at 115 MeV/nucleon and comparison with results obtained from candidate T=2 states in the present $^{26}$Mg($^{3}$He,t) experiment at 140 MeV/nucleon and two (d,$^{2}$He) experiments \cite{NII94,XU96}}
\begin{ruledtabular}
\begin{tabular}{cccccccc}
\multicolumn{2}{c}{(t,$^{3}$He) present data} &\multicolumn{2}{c}{($^{3}$He,t) present data \footnotemark[1]} &\multicolumn{2}{c}{(d,$^{2}$He) \footnotemark[2]} &\multicolumn{2}{c}{(d,$^{2}$He)\footnotemark[3]} \\ \hline
E$_{x}$ (MeV)& B(GT) & E$_{x}$ (MeV) & B(GT)          & E$_{x}$ (MeV) & B(GT)          & E$_{x}$ (MeV) & B(GT) \\ \hline 
0.08\footnotemark[4] & 0.41(5)$\pm$0.03 & 13.57&  0.41$\pm$0.02 & 0.08\footnotemark[4] & 0.38 & 0.08\footnotemark[4] & 0.44$\pm$0.04\\
     &               & 14.53&  0.09$\pm$0.02 &      &&&\\ 
1.4$\pm$0.2 & 0.09$\pm$0.02  & 14.88&  0.11$\pm$0.03 & 1.5 & 0.06 &&\\
2.6$\pm$0.2 & 0.13$\pm$0.02  & 15.91&  0.17$\pm$0.03 & 2.6 & 0.11 &&\\
5.1$\pm$0.4 & 0.22$\pm$0.04  & 18.32&  0.13$\pm$0.06 & 5.2 & 0.09 &&\\ \hline
Sum         & 0.85$\pm$0.06  &      &  0.91$\pm$0.09(0.82$\pm$0.08\footnotemark[5]) & & 0.64 \\
\end{tabular}
\end{ruledtabular}
\footnotetext[1]{Potential T=2 states, assuming isospin symmetry and multiplying with a factor of 6 because of a difference in Clebsch-Gordan coefficients.}
\footnotetext[2]{From Ref. \cite{NII94}}
\footnotetext[3]{From Ref. \cite{XU96}}
\footnotetext[4]{Excitation energy fixed to known location of $1^{+}$ state \cite{PEA87}.}
\footnotetext[5]{Value excluding the state at 14.53 MeV.}
\end{table*}

\subsection{The Gamow-Teller sum rule and the comparison of experiment with shell-model theory} 
For the Gamow-Teller strength in nuclei, a model-independent sum-rule can be derived \cite{IKE63}:
\begin{equation}
\label{eq:sumrule}
S_{\beta^{-}}(GT)-S_{\beta^{+}}(GT)=3(N-Z)
\end{equation} 
Experimentally, only about 50-60\% of the sum-rule strength is observed at excitation energies below $\sim$20 MeV \cite{GAA81,GAA85} (see also Section \ref{sec:dwba}). Different mechanisms have been proposed to explain this quenching phenomenon. One explanation is that due to mixing with 2$p$-2$h$ configurations via the strong tensor interaction, Gamow-Teller strength is pushed up in excitation energy \cite{HYU80,ARI99}. Experimentally, such high-lying strength has been found in the study of $^{90}$Zr(p,n) at 295 MeV \cite{WAK97} and $^{90}$Zr(n,p) \cite{RAY90} and \cite{SAK04,YAK05}. By combining these data, the measured sum-rule strength is 88$\pm$6\% of $3(N-Z)$, not including the uncertainty of the Gamow-Teller unit cross section of 16\% \cite{SAK04}. This result indicates that a significant fraction of the missing Gamow-Teller strength is pushed up in excitation energy due to configuration mixing. The second mechanism suggested to explain quenching is that, due to coupling between the $\Delta$(1232)-isobar nucleon-hole state and the 1$p$-1$h$ Gamow-Teller state, strength would shift to excitation energies of about 300 MeV \cite{ERI73}. Of course, the two mechanisms are not mutually exclusive. In fact, an analysis of isoscalar magnetic moments and Gamow-Teller matrix elements in the sd-shell suggest that the quenching factor is $Q=(1-\delta_{cm}-\delta_{\Delta})^2=0.61$ where $\delta_{cm}=0.15$ comes from configuration mixing and $\delta_{\Delta}=0.07$ comes from $\Delta(1232)$-isobar nucleon-hole admixtures \cite{BRO88}. 

In the present shell-model calculations neither of the above quenching mechanisms is taken into account. In the experiments discussed here, Gamow-Teller strength in the continuum could not be extracted, since it requires an MDA using a large number of multipoles, fitted over a wider scattering-angle range. We should, therefore, expect to see only a fraction of the full sum rule strength. Indeed, by combining the $^{26}$Mg($^{3}$He,t) and $^{26}$Mg(t,$^{3}$He) results, we find:
\begin{eqnarray}
\label{eq:sumrule2}
S_{\beta^{-}}(GT)-S_{\beta^{+}}(GT)&=&(4.46\pm0.05-0.85\pm0.06)\nonumber\\
&=&3.61\pm0.08,
\end{eqnarray}
which corresponds to 60\%$\pm$1\% (the error is due to statistical and systematical uncertainties in the background estimates only) of the $3(N-Z)=6$ sum rule. This agrees well to the established quenching factor in the sd-shell  \cite{WIL83,BRO85} of 0.59$\pm$0.03 and with quenching factors observed in this mass region in studies using the (p,n) reaction \cite{GAA85}. The orbital dependence of the Gamow-Teller effective operator in \cite{BRO85} is small and the Gamow-Teller strength spectrum  is essentially the same as that obtained with an overall quenching factor of 0.59

\begin{figure*}
\includegraphics[scale=0.85]{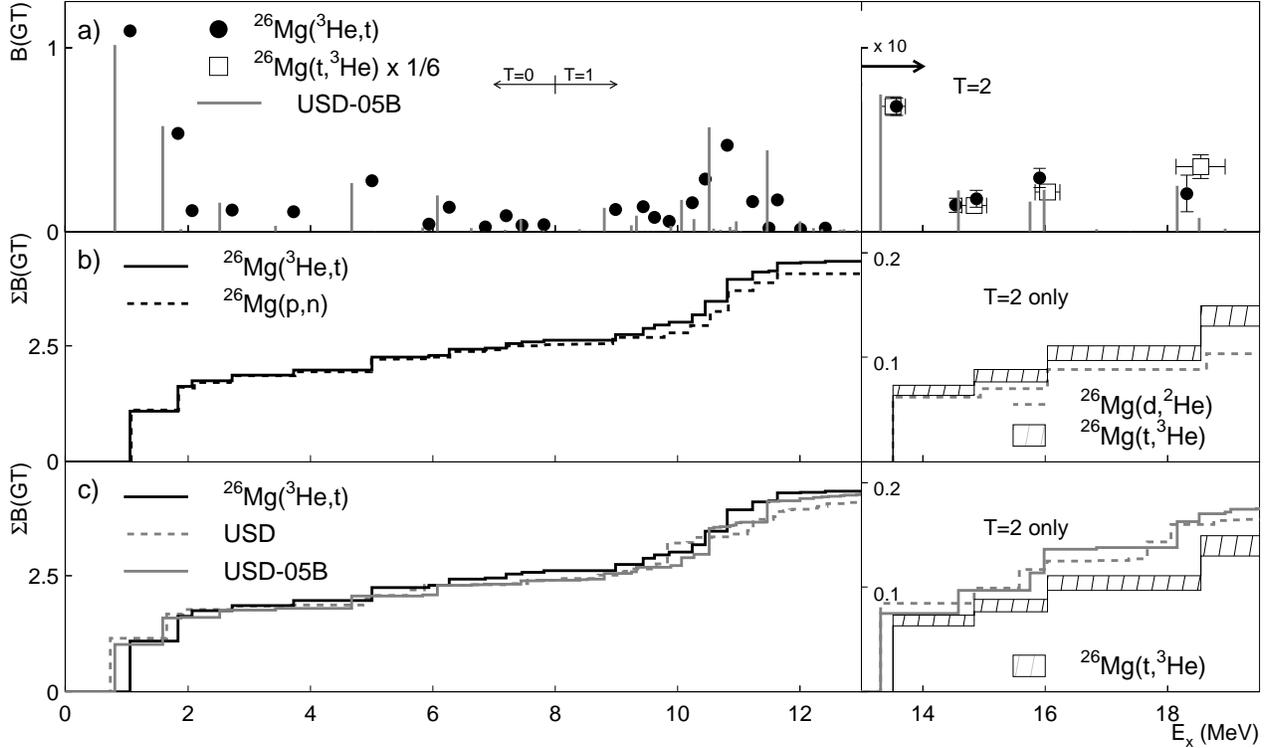}
\caption[]{\label{prc5}a) Measured Gamow-Teller strengths from  $^{26}$Mg($^{3}$He,t) and comparison with the shell-model predictions using the USD-05B interaction, multiplied by the extracted quenching factor of 0.6. For $E_{x}>13$ MeV, Gamow-Teller strengths extracted from  $^{26}$Mg(t,$^{3}$He) are included. The excitation energies in $^{26}$Na have been shifted as explained in Fig. \ref{prc3} and the strengths have been divided by 6 to account for the difference in isospin Clebsch-Gordan coefficients. Regions where T = 0, 1 or 2 are dominant are indicated. b) Cumulative sums of Gamow-Teller strengths; below $E_{x}=13$ MeV, results from $^{26}$Mg($^{3}$He,t), $^{26}$Mg(p,n) are compared. Above $E_{x}=13$ MeV, results from $^{26}$Mg(t,$^{3}$He) and $^{26}$Mg(d,$^{2}$He) \cite{NII94} experiments are compared. Note the changes in vertical scale for $E_{x}>13$ MeV. c) Comparison between results from the $^{26}$Mg($^{3}$He,t) ($E_{x}<13$ MeV) and $^{26}$Mg(t,$^{3}$He) ($E_{x}>13$ MeV) experiments and the shell-model calculations using the USD and USD-05B interactions. }
\end{figure*}

Next, a more detailed comparison between the shell-model calculations and the data is made. The results are summarized in Fig. \ref{prc5}. In Fig. \ref{prc5}a, the extracted Gamow-Teller strengths from the $^{26}$Mg($^{3}$He,t) and $^{26}$Mg(t,$^{3}$He) experiments are compared with the calculation performed with the updated sd-shell interaction, USD-05B \cite{USDN}. In this interaction, the additional experimental information that has been collected on binding and excitation energies in the sd-shell since the development of the original USD \cite{USD} interaction in 1983, is included in the fit of the parameters in the interaction. The theoretical Gamow-Teller strengths shown in Fig. \ref{prc5} were calculated using the code OXBASH \cite{OXBA} in the full sd shell-model space. These Gamow-Teller strengths were then multiplied with 0.6 to account for the quenching phenomenon described above. 

At $E_{x}>13$ MeV, the shell-model calculations are compared with both $^{26}$Mg($^{3}$He,t) and $^{26}$Mg(t,$^{3}$He) data sets in Fig. \ref{prc5}a. The regions where transitions with certain isospin (T=0,1, or 2) are expected to be dominant from theory are indicated.

The general correspondence of strength distributions is facilitated by comparing the cumulative summed strengths.
In Fig. \ref{prc5}b, the extracted cumulative sums of Gamow-Teller strengths of transitions in $^{26}$Mg($^{3}$He,t) (below $E_{x}$=13 MeV) and $^{26}$Mg(t,$^{3}$He) (above $E_{x}$=13 MeV) are compared with the experimental results from $^{26}$Mg(p,n) \cite{MAD87} and $^{26}$Mg(d,$^{2}$He) \cite{NII94}, respectively. In spite of minor differences, discussed above, there is an overall good agreement between results from the different probes.

In Fig. \ref{prc5}c, the $^{26}$Mg($^{3}$He,t) (below $E_{x}$=13 MeV) and $^{26}$Mg(t,$^{3}$He) (above $E_{x}$=13 MeV) are compared with shell-model calculations. Besides the theoretical results using the new USD-05B interaction \cite{USDN}, the ones using the old USD interaction \cite{USD} are included as well. The USD and USD-05B interactions both give Gamow-Teller strength distributions in good overall agreement with experiment. The USD-05B interaction is better for the T=1 states in the region 8-10 MeV.
With the USD interaction, the strength distribution is shifted to slightly lower energies and somewhat less strength is predicted than with the USD-05B interaction. Using the updated interaction results in a better agreement with the data. In the T=2 region (above $E_{x}$=13 MeV), both calculations predict slightly higher strengths than the data. The shapes of the experimental and theoretical distributions are similar. 

\section{The DWBA and systematic uncertainties in the extraction of Gamow-Teller strength}
\label{sec:dwba}
In this section, the DWBA calculations used in Section \ref{sec:strength} are described in detail. It was shown in Fig. \ref{prc4} that the calculations reproduce the measured angular distributions well. Here we will compare the absolute cross sections of the data and the theory. We will also give a theoretical estimate for the main source of systematic errors in the extracted Gamow-Teller strengths due to interference between $\Delta L=0$ and $\Delta L=2$ (both $\Delta S=1$) transitions that lead to $1^{+}$ states.

\subsection{Details of the DWBA calculations}
For many ($^{3}$He,t) experiments carried out in the past, DWBA calculations have been performed using the code DW81 \cite{DW81} and an effective $^{3}$He-nucleon interaction \cite{SCH71,WER90,GRA86} consisting of 4 components, associated with the $\sigma\tau$ (spin and isospin transfer), $\tau$ (isospin transfer), $LS\tau$ (spin-orbit) and tensor-$\tau$ interactions. The $\sigma\tau$ and $\tau$ terms are represented by a single Yukawa potential, with a fixed range ($R=1.414$ fm) set equal to the range of the one-pion exchange potential (OPEP). It was argued that, when averaged over the volume of the $^{3}$He particle, components of shorter range, arising from $\rho$-meson or $2\pi$-exchange at the nucleon-nucleon level, result in ranges close to the OPEP value. The tensor term is represented by a potential of the form $r^{2}\times\text{Yukawa}$ (with range $R=0.878$ fm), multiplied by the tensor operator $S_{12}$. The spin-orbit term is usually set to zero, based on analyses of angular distributions for excitation of the isobaric analog states. The amplitudes of the $\sigma\tau$, $\tau$ and tensor-$\tau$ terms are then chosen to reproduce experimental data. The uncertainty of the tensor-$\tau$ term is large since there are very few transitions in which this component can be isolated. 
Note that exchange contributions to the calculated cross sections with this interaction are absorbed in the effective $^{3}$He-nucleon interaction and not separable from the direct contributions. 

Although the calculations with the effective $^{3}$He-nucleon interaction generally lead to a good match with the data at forward scattering angles, its use is somewhat unsatisfactory since little information can be extracted about the reaction mechanism and thus about effects that possibly lead to uncertainties in extraction of Gamow-Teller strengths. An important question is whether it is possible to construct this effective interaction from effective nucleon-nucleon potentials used successfully in analyses of (p,n) and (n,p) data. This requires the folding of such interactions over the transition densities and taking into account the finite sizes of the $^{3}$He and triton particles as well as exchange contributions.  

We, therefore, used the code FOLD \cite{FOLD}, in which the Love-Franey nucleon-nucleon interaction \cite{LOV81,LOV85}, briefly described in section \ref{sec:prel} and commonly used in the analysis of (p,n) and (n,p) data, is double-folded over the transition densities. Central, spin-orbit and tensor components of the interaction are included and exchange is treated in the short-range approximation described in Ref. \cite{LOV81}. The exact treatment of exchange for composite particles is a challenge, and in the case of charge-exchange reactions, has only been attempted at beam energies of 200 and 300 MeV/nucleon \cite{UDA87,KIM00}. Small changes to the code FOLD were made so that Clebsch-Gordan coefficients were properly determined and exchange terms were calculated according to the procedure described in Ref. \cite{LOV81}. Tensor exchange effects are ignored, which is a reasonable approximation, as shown in Ref. \cite{UDA87}. In this reference it was also shown that the no-recoil approximation for exchange, which is in effect similar to the short-range approximation used in the present paper \cite{PET69}, results in an underestimation of exchange effects. Since direct and exchange contributions interfere destructively, cross sections should be overestimated.
It was shown, however, that the angular distributions hardly changed if exchange was treated in the approximate manner instead of using an exact finite range calculation \cite{UDA87}.
 
One-body transition densities (OBTDs) were calculated with the code OXBASH \cite{OXBA} employing the new USD-05B interaction \cite{USDN} in the sd shell-model space. Radial wave functions of the target and residue were calculated using a Woods-Saxon potential. Binding energies of the particles were determined in OXBASH \cite{OXBA} using the Skyrme SK20 interaction \cite{BRO98}. They are -7.74 MeV (-10.00 MeV), -3.69 MeV (-5.27 MeV), -10.65 MeV (-12.12 MeV) for the protons (neutrons) in the $s_{1/2}$, $d_{3/2}$ and $d_{5/2}$ orbits of $^{26}$Mg, respectively; -6.00 MeV, -1.67 MeV, and -8.47 MeV for the protons in the $s_{1/2}$, $d_{3/2}$ and $d_{5/2}$ orbits of $^{26}$Al and -7.54 MeV, -10.65 MeV and -3.911 MeV for the neutrons in the $s_{1/2}$, $d_{3/2}$, and $d_{5/2}$ orbits of $^{26}$Na. For $^{3}$He and $^{3}$H, densities were obtained from Variational Monte-Carlo results \cite{WIR05}. 

Optical potential parameters (OPP) were obtained by refitting $^{3}$He elastic-scattering data on $^{28}$Si \cite{YAM95} at 150 MeV/nucleon because of the large discrepancies found between OPP for other nuclei in Refs. \cite{YAM95} and \cite{KAM03}. Following the definitions of the optical potentials in these references, the new parameters are: $V_{R}=-25.1$ MeV, $r_{R}=1.43$ fm, $a_{R}=0.833$ fm, $W_{I}=-40.0$ MeV, $r_{I}=0.963$ fm, $a_{I}=1.03$ fm and $r_{C}=1.25$ fm. With the old parameters, angular distributions calculated in DWBA for the ($^{3}$He,t) and (t,$^{3}$He) reactions did not match the data well. With the refitted parameters (see Fig. \ref{prc4}) the agreement is good. Following Ref. \cite{WER89}, the depths of the triton potentials were calculated by multiplying the depths of the $^{3}$He potentials by 0.85, while leaving radii and diffusenesses constant.

\subsection{Comparison of absolute cross sections in experiment and theory.}
In Table \ref{tab:table3}, a comparison is made between measured and calculated cross sections at 0$^{\circ}$ for the four Gamow-Teller transitions to $1^{+}$ states in $^{26}$Al for which the Gamow-Teller strength is known from $\beta$-decay, as well as the excitation of the IAS.
The excitation of the IAS exhausts the full Fermi sum rule strength (N-Z)=2. The comparison between the experimental result and the theoretical calculation for this transition, therefore, has to be made without corrections. For the Gamow-Teller transitions, one has to take into account that the strengths calculated in the shell-model are not equal to the measured strengths in $\beta$-decay data due to the quenching of the Gamow-Teller strengths and minor additional differences between measured and calculated strengths on a state-by-state basis. In addition a small multiplicative correction (see footnote in Table \ref{tab:table3}) has to be made due to the fact that the Gamow-Teller strength for the transition from $^{3}$He to the triton is $2.685\pm0.004$ \cite{CHO93} instead of 3 as assumed in the calculations.  

\begin{table*}
\caption{\label{tab:table3} Comparison between measured and calculated differential cross sections at $0^{\circ}$ for the excitation of the IAS and the first four $1^{+}$ states in $^{26}$Al via the ($^{3}$He,t) reaction. The calculated cross sections for Gamow-Teller transitions are corrected for reasons discussed in the text (see footnote for equation used), before comparing them with the data. }
\begin{ruledtabular}
\begin{tabular}{ccccccccc}
\multicolumn{2}{c}{state in $^{26}$Al} & \multicolumn{3}{c} {Strength} & \multicolumn{3}{c}{$\frac{d\sigma}{d\Omega}(0^{\circ})$ (mb/sr) }& ratio \\ 
\cline{1-2} \cline{3-5} \cline{6-8} \\
E$_{x}$ (MeV) & $J^{\pi}$ & B$_{\text{USD-05B}}$ & B$_{\beta-\text{decay}}$ & B$_{(^{3}\text{He},t)}$ & DWBA & DWBA corrected\footnotemark[1] & experiment &  $\frac{\text{DWBA corrected}}{\text{experiment}}$ \\ \hline
0.228 &  $0^{+}$ (IAS) & 2\footnotemark[2] & -                & 2\footnotemark[2] & 7.20 &7.20 & 4.0$\pm$0.1  & 1.80$\pm$0.05 \\
1.06  &  $1^{+}$ (GT)  & 1.69              & 1.098$\pm$0.022  & 1.09$\pm$0.03     & 38.9 &22.6 & 13.9$\pm$0.3 & 1.62$\pm$0.04 \\
1.85  &  $1^{+}$ (GT)  & 0.96              & 0.536$\pm$0.014  & 0.54$\pm$0.02     & 21.5 &10.7 & 6.7$\pm$0.2  &
1.60$\pm$0.05 \\
2.07  &  $1^{+}$ (GT)  & 0.022             & 0.091$\pm$0.004  & 0.114$\pm$0.008   & 0.56 &2.08 & 1.45$\pm$0.03&
1.43$\pm$0.03 \\
2.74  &  $1^{+}$  (GT)  & 0.26              & 0.113$\pm$0.005  & 0.119$\pm$0.008   & 5.06 &1.97 & 1.5$\pm0.03$ & 1.31$\pm$0.03 \\
\end{tabular}
\end{ruledtabular}
\footnotetext[1]{No correction is applied for the IAS. For GT transitions, $\frac{d\sigma}{d\Omega}(0^{\circ})_{\text{corrected}}=\frac{d\sigma}{d\Omega}(0^{\circ})_{\text{DWBA}}\frac{B(GT)_{\beta}}{B(GT)_{USD-05B}}\frac{B(GT)_{(t,^{3}He)}}{3}$}
\footnotetext[2]{Using the Fermi sum rule  $S_{-}(F)-S_{+}(F)=(N-Z)$ with $S_{+}=0$.}
\end{table*}

Ignoring the transitions to the states at $E_{x}$=2.07 MeV and $E_{x}$=2.74 MeV for reasons discussed below, the differential cross sections calculated via DWBA overestimate the data by factors of 1.80$\pm$0.05 for the transition to the IAS and 1.62$\pm$0.04 and 1.60$\pm$0.05 for the transitions to the first two $1^{+}$ states at $E_{x}$=1.06 MeV and $E_{x}$=1.85 MeV, respectively. Uncertainties in the beam integration and target thicknesses ($<10\%$) and optical potential parameters ($\sim20\%$) cannot account for this discrepancy. As mentioned above, one of the likely sources is the approximate treatment of the exchange processes \cite{UDA87}. However, even when such processes were treated exactly, a significant overestimation ($\sim 45$\%) was found \cite{UDA87}. The authors of Ref. \cite{UDA87} partially attributed this to the optical potential parameters used, and in Ref. \cite{KIM00} the strength of the imaginary potential (in that case the potentials were calculated in a single-folding procedure) was reduced  by 30\%. Since we used optical model parameters obtained from fitting elastic scattering data, and since a drastic change of the strength of the imaginary potential would result in a bad description of the measured angular distributions, such a change can not be justified in the analysis of the current data. It can not be excluded that other effects, such as a density-dependence of the interaction, play a role as well. In fact, for the (p,n) reaction, corrections of up to 40\%, and a better fit to the data as a result, have been reported \cite{CHE92} if density-dependence effects were taken into account. Changes due to the density-dependence of the interaction are expected to be stronger for Fermi transitions than for Gamow-Teller transitions, since the potentials involved have shorter ranges \cite{LOV87}. Since exchange effects are different for Fermi and Gamow-Teller transitions as well, it is not possible to draw strong conclusions based on the nature of small differences in the discrepancies between data and DWBA calculations for the excitation of the IAS and the Gamow-Teller transitions. It indicates, however, that caution is advised when calibrating the unit cross section for Gamow-Teller transitions using the ratio to the unit cross section for Fermi transitions, as is sometimes done.   

We thus find that if the Love-Franey interaction \cite{LOV81,LOV85} is used in Eq. \ref{eq:eik} with the approximate treatment of exchange, an additional factor $C$ is to be included: 
\begin{equation}
\label{eq:eikc}
\frac{d\sigma}{d\Omega}(q=0)=CKN|J_{\sigma\tau}|^{2}B(GT),
\end{equation}
with $C=0.62$ for Gamow-Teller transitions and $C=0.56$ for Fermi transitions.

It is not well-known whether discrepancies between data and the DWBA calculations are present for other composite probes as well. In Ref. \cite{GRE04} a scaling factor $C=0.32$ is determined for the (d,$^{2}$He) reaction from a comparison to transitions with known B(GT). This factor includes a correction for the cut on the relative energy between the two outgoing protons (internal energy $\epsilon_{pp}<1$ MeV), which is hard to estimate theoretically. It is, therefore, not clear whether the cut on the relative energy between the protons fully explains the value of $C$ in the (d,$^{2}$He) reaction, or whether additional contributions, like corrections to the treatment of exchange, are important as well. If the (d,$^{2}$He)  reaction is considered to be nearly quasifree, with the projectile proton acting as a spectator, the mechanisms involved could be quite different than for the (t,$^{3}$He) reaction.

\subsection{Uncertainties in the proportionality between Gamow-Teller strength and cross section at zero momentum transfer} 
As shown in Table \ref{tab:table3}, the discrepancies between the DWBA calculations and the data are different for the two Gamow-Teller transitions to the lowest-lying $1^{+}$ states at $E_{x}=1.06$ MeV and $E_{x}=1.85$ MeV) and the transitions to the next two $1^{+}$ states at $E_{x}=2.07$ MeV and $E_{x}=2.74$ MeV. This constitutes a breaking of the proportionality described in Eq. \ref{eq:eik}. Next, we discuss why the calibration of the proportionality was performed using the first two transitions only. It will be shown that the magnitude of the proportionality breaking can be explained as being due to interference between $\Delta L=2$, $\Delta S=1$ amplitudes (mediated via the tensor-$\tau$ component of the interaction) and $\Delta L=0$, $\Delta S=1$ amplitudes (mediated via the $\sigma\tau$ component of the interaction). It is important to note that, in practice, it is impossible to make corrections to Gamow-Teller strengths deduced from charge-exchange reactions for this type of proportionality breaking, since it requires a-priori and exact knowledge of the wave functions.

It is well-known \cite{TAD87} that proportionality breaking occurs, and stronger so for transitions with small Gamow-Teller strengths. 
To quantify the proportionality breaking for the ($^{3}$He,t) and (t,$^{3}$He) reactions at the beam energies used for the experiments described here, a theoretical study was performed as shown in the following. Similar studies have been performed for the (p,n) reaction (see e.g. \cite{AUS94}). 

OBTDs for the first 100 transitions to T=0, T=1 and T=2 $1^{+}$ states (300 states in total) in $^{26}$Al via $^{26}$Mg($^{3}$He,t) at 140 MeV/nucleon were generated in the shell-model code OXBASH \cite{OXBA}, using the USD-05B \cite{USDN} interaction. The Gamow-Teller strength and subsequently the differential cross sections in DWBA are calculated for each of the transitions, as described above. From this point onward, we treated these calculated cross sections as if they were data and extracted the Gamow-Teller strengths for each state in exactly the same manner as was done in the analysis of the experimental data described in section \ref{sec:strength}. However, in the calculated DWBA cross sections it is possible to switch on and off the interference due to $\Delta L=2$, $\Delta S=1$ contributions. In principle, this should be done on the level of the OBTDs. It was found, however, that by simply setting the tensor-$\tau$ component of the interaction equal to zero, the same effect was achieved: the proportionality between cross section at zero-momentum transfer and Gamow-Teller strength was near perfect and remained within deviations of $\sim2$\%). This indicates that the tensor-$\tau$ component of the interaction that mediates the $\Delta L=2$, $\Delta S=1$ transitions, is indeed the dominant cause for the proportionality breaking. Finally, we determined the difference in extracted Gamow-Teller strength for each state from the calculations where the interference was allowed to occur (as in the real data) and the input Gamow-Teller strength (as calculated in the shell model). The following relative systematic error was defined for each transition:
\begin{equation}
\label{eq:syst}
\text{Rel. syst. error}=\frac{B(GT)_{\text{DWBA}}-B(GT)_{\text{SM}}}{B(GT)_{\text{SM}}},
\end{equation}         
where $B(GT)_{\text{DWBA}}$ is the Gamow-Teller strength extracted via the calculations in DWBA with the full interaction (i.e. the simulated data with the tensor-$\tau$ interaction) and $B(GT)_{\text{SM}}$ is the Gamow-Teller strength of the corresponding transition as calculated in the shell model. 

\begin{figure}
\includegraphics[width=8.5cm]{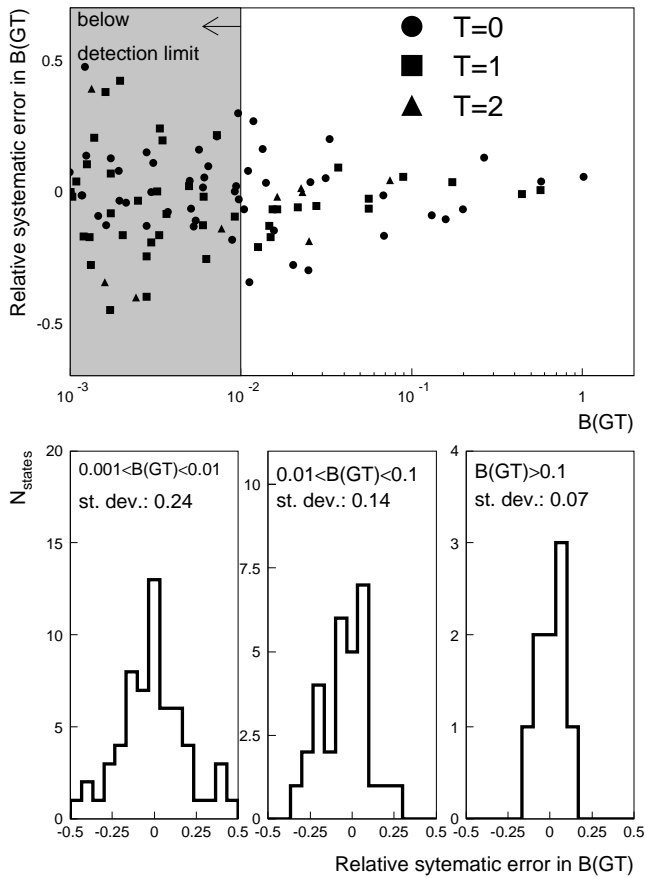}
\caption{\label{prc6} Results of a theoretical study into the effects of $\Delta L=2$, $\Delta S=1$ contributions mediated via the tensor-$\tau$ interaction that interfere with $\Delta L=0$, $\Delta S=1$ contributions to Gamow-Teller transitions. The calculations were performed for the $^{26}$Mg($^{3}$He,t) reaction at 140 MeV/nucleon and included all Gamow-Teller transitions to T=0, T=1 and T=2 $1^{+}$ states generated in the shell-model calculations described in the text up to an excitation energy of 20 MeV in $^{26}$Al. Only transitions with B(GT)$>0.001$ are shown. In the top panel the expected relative error made in the extraction of the Gamow-Teller strength is shown as a function of the shell-model strength. The histograms in the bottom panel contain the relative errors for transitions with $0.001<\text{B(GT)}<0.01$ (left), $0.01<\text{B(GT)}<0.1$ (center) and $\text{B(GT)}>0.1$ (right). Standard deviations of the distributions are indicated.
}
\end{figure}

In Fig. \ref{prc6}, the results from the theoretical study are displayed. In the top panel, the relative systematic errors calculated via Eq. \ref{eq:syst} are plotted against Gamow-Teller strengths generated in the shell-model (note the logarithmic horizontal scale). These shell-model strengths were multiplied by the experimentally establish quenching factor of 0.6, in order to enable a direct comparison with strengths deduced from experiments. Transitions with Gamow-Teller strengths smaller than 0.001 are excluded from the plot and the approximate experimental detection limit is indicated. It is clear that for large strengths, the systematic errors are small. They increase with decreased strength. No clear dependence on isospin of the final state was found.
In the bottom three panels, the relative systematic errors are plotted for three regions with increasing strengths. In each panel, the standard deviation of the relative systematic error is given, decreasing from 0.24 for B(GT)$<0.01$ to 0.14 for 0.01$<$B(GT)$<$0.1 and to 0.07 for B(GT)$>$0.1. Note that the interference can be destructive or constructive, and that, on average the systematic error is close to zero. In fact, when integrated over all Gamow-Teller transitions, the extracted strength in the theoretical study above only deviated by about 1\% from the integrated Gamow-Teller strength calculated in the shell model. It was found that the standard deviation of the relative systematic error can be approximated with:
\begin{equation}
\label{eq:syst2}
\sigma_{\text{rel. syst. error}}\approx 0.03-0.035\ln(\text{B(GT)}).
\end{equation}

From this study, it is clear that calibration of the proportionality is best performed using transitions with large Gamow-Teller strengths, hence the choice made earlier to only use the first two strong Gamow-Teller transitions for this purpose. For the states at $E_{x}=2.07$ MeV (B(GT)=0.091) and at $E_{x}=2.74$ MeV (B(GT)=0.113), standard deviations of 11.4\% and 10.6\% are expected, respectively, using Eq. \ref{eq:syst2}. The measured deviations, assuming that the proportionality calibration using the transitions to the states at $E_{x}$=1.06 MeV and $E_{x}$=1.85 MeV is perfect, are 10.6\% (0.93 standard deviation) and 18.5\% (1.74 standard deviations), respectively. Although the sample size is small, this result provides a good indication that the proportionality breaking can be explained as being mainly due to the interference from $\Delta L=2$, $\Delta S=1$ components in the Gamow-Teller transitions.    
Note that the proportionality breaking due to this phenomenon is expected to be different for reactions that strongly probe the nuclear surface (like ($^{3}$He,t) and (t,$^{3}$He)) than reactions that penetrate the nucleus deeply, like (p,n) and (n,p). For the $^{26}$Mg($^{3}$He,t) and $^{26}$Mg(t,$^{3}$He) reactions discussed in this paper, setting the lower boundary of the radial integral over the form factor to 1.9 fm (the radius of $^{26}$Mg is approximately 3.8 fm) only reduces the calculated cross sections in DWBA by about 5\%.

\section{Conclusions and Outlook}
\label{sec:conclusions}
In summary, we have shown that the (t,$^{3}$He) reaction at 115 MeV/nucleon is an accurate probe for Gamow-Teller strength in the $\Delta$T$_{z}=+1$ direction. In addition to (n,p) and (d,$^{2}$He), it provides a new tool to test weak rates used in stellar-evolution modeling. Together with ($^{3}$He,t), (t,$^{3}$He) provides an alternative to (p,n) and (n,p) reactions with significantly improved resolution.
The ability to predict the Gamow-Teller strength distribution with the new USD-05B interaction was tested and a good correspondence with the experiment was found. Compared to the old USD interaction, improvements were most significant for transitions to states with isospin T=1.

Except for a general scaling factor, differential cross sections for a variety of $J^{\pi}$ transitions are reproduced in double-folding DWBA using the code FOLD, and systematic errors on the level of 10-20\%in the extraction of Gamow-Teller strengths on a transition-by-transition basis can be understood in terms of interference between the $\Delta L=0$, $\Delta S=1$ and $\Delta L=2$, $\Delta S=1$ amplitudes. When integrated over many states, the systematic error becomes small ($\sim1\%$), because the deviations for various states can be positive or negative and cancel. Understanding the discrepancy between measured and theoretical absolute cross sections by a factor of about 1.6-1.8 for Gamow-Teller and Fermi transitions will require further investigation; a likely cause is the approximate treatment of exchange in the DWBA calculations. A study of such issues on nuclei over a wide mass range is in progress and will be presented in a forthcoming paper.

Increased triton beam intensities have recently been obtained at the NSCL, using a primary $^{16}$O beam instead of a primary $\alpha$-beam \cite{HIT05}. This will provide 5-10 times better statistics. Meanwhile, beam-tuning capabilities have improved as well which makes it easier to achieve optimum dispersion-matching conditions. Somewhat thinner targets than the one used for the $^{26}$Mg(t,$^{3}$He) can be used as well, so that resolutions of about 200 keV can reliably be achieved. The improved capabilities will enable measurements of weak-interaction strengths using targets of importance for understanding the late stages of stellar evolution.

\begin{acknowledgments}
We thank the cyclotron staffs at the NSCL and RCNP for their support during the experiments described in this paper.
This work was supported by the US NSF (PHY02-16783 (JINA), PHY-0110253 and PHY-0244453), the Ministry of Education, Science, Sports and Culture of Japan, the Stichting voor Fundamenteel Onderzoek der Materie (FOM), the Netherlands and by the Office of the Vice President for Research, University of Michigan.
\end{acknowledgments}

\bibliography{prc}

\end{document}